\documentclass[twocolumn]{revtex4}
\usepackage{amsmath}
\usepackage{amssymb}
\usepackage{latexsym}
\usepackage{epsfig}
\usepackage{color}
\begin{document}

\title{Minimal length: A source of quantum non-locality}
\author{H. Moradpour$^{1}$\footnote{h.moradpour@riaam.ac.ir}, S. Jalalzadeh$^{2,3,4}$\footnote{shahramjalalzadeh@iyte.edu.tr}}
\address{$^1$ Research Institute for Astronomy and Astrophysics of Maragha
(RIAAM), P.O. Box 55134-441, Maragha, Iran\\
$^2$ Izmir Institute of Technology, Department of Physics, Urla,
35430, Izmir, T\"{u}rkiye\\
$^3$ Center for Theoretical Physics, Khazar University, 41
Mehseti Street, Baku, AZ1096, Azerbaijan\\
$^4$ Department of Physics, Dogus University,
Dudullu-\"{U}mraniye, 34775 Istanbul, T\"{u}rkiye}

\begin{abstract}
The narrow and subtle difference between the Hilbert spaces of
operators corresponding to the canonical momentum and the
generalized momentum that includes minimal length effects is
polished. Consequently, complex eigenvalues may be allowed for the
canonical momentum operator due to the existence of minimal
length. A novel quantum entanglement generation is also reported
indicating the power of theories including a minimal length in
enriching the current understanding of quantum non-locality.
\end{abstract}

%\date{\today}
%provided that a measurement of canonical momentum confirms them
\maketitle

Based on the Heisenberg Uncertainty Principle (HUP), there is no
limit on the minimum value of length, provided that the momentum
information is completely lost. On the other hand, the role of
gravity at quantum scale is mysterious
\cite{Hossenfelder:2012jw,Das:2008kaa,Ali:2009zq,Ali:2011fa}.
Nevertheless, considering Newtonian gravity in revisiting the
Heisenberg microscope unveils a non-zero minimal length
\cite{Hossenfelder:2012jw,Mead:1964zz,Mead:1966zz}. It leads to
the Generalized Uncertainty Principle (GUP), which is also
reinforced in quantum scenarios of gravity
\cite{Hossenfelder:2012jw,Kempf:1994su}. HUP lies at the heart of
classically amazing phenomena like quantum non-locality
\cite{Einstein:1935rr} and correspondingly, the existence of
minimal length and quantum aspects of gravity affect the quantum
entanglement, the violation of Bell's inequality
\cite{Bose:2017nin,Marletto:2017kzi,Blado:2017gnf,Aghababaei:2021yzx,Aghababaei:2022rqi,Park:2022jrs,Filho:2023vpx,Marletto:2025fpm,Moradpour:2023gsn},
and the information bounds
\cite{Pedram:2016gps,Rastegin:2016xeb,Aghababaei:2022jxd}. Indeed,
we are even witnessing a fascinating claim that classical gravity
produces quantum entanglement
\cite{Marletto:2025fpm,Aziz:2025ypo}. Additionally, the
conjectured correspondence between metric-affine gravity and
quantum gravity scenarios \cite{Ali:2024tbd} further motivates us
to study the implications of the minimum length on various
phenomena such as quantum non-locality and related topics.

In Quantum Mechanics (QM), momentum ($\hat{p}$) and position
operator ($\hat{x}$) are conjugate and thus Fourier transform
dual. They do not commute with each other (the origin of the EPR
paradox), and it is enough to replace momentum and position in a
classical Hamiltonian with the corresponding operators to get the
quantum counterpart of the system. Usually, the position
representation is taken into account to write the Schr\"{o}dinger
equation for which $\hat{p}=-i\hbar\nabla$ \cite{Brau:1999uv}. In
this manner, the effects of minimal length are gathered in the
change of momentum operator from canonical momentum ($\hat{p}$) to
generalized momentum ($\hat{P}$) as
$\hat{p}\rightarrow\hat{P}=\hat{p}~(1+\beta\hat{p}^2)$, up to the
first order of the GUP parameter $\beta$
\cite{Hossenfelder:2012jw,Kempf:1994su,Brau:1999uv}.
Correspondingly, to study the implications of minimal length on a
quantum mechanical system with Hamiltonian $\hat{H}$, it is enough
to replace $\hat{p}$ with $\hat{P}$
\cite{Hossenfelder:2012jw,Kempf:1994su}. Briefly, while $\hat{P}$
is claimed to be the physical momentum operator in the presence of
minimal length, canonical momentum $\hat{p}$ is still an operator
with its eigenvalues and Hilbert space.

The EPR intelligently experiment \cite{Einstein:1935rr} reveals
the non-local nature of QM that finally opens a new window towards
information science by relying on quantum non-locality capacity,
originally rooted in HUP
\cite{Einstein:1935rr,Bohm:1957zz,Heisenberg,Franson,oppen,Peres:2002ip}.
Despite previous works on $i$) testing the quantum nature of
gravity using entangled states
\cite{Bose:2017nin,Marletto:2017kzi,Altamirano:2016fas,Hall:2017nzl,Doner:2022npz,Spaventa:2023eys}
and $ii$) the effects of GUP on Bell inequality, quantum
non-locality, and thus information bounds
\cite{Blado:2017gnf,Aghababaei:2021yzx,Aghababaei:2022rqi,Park:2022jrs,Filho:2023vpx,Moradpour:2023gsn,Pedram:2016gps,Rastegin:2016xeb,Aghababaei:2022jxd},
the role and ability of minimal length (GUP or quantum aspects of
gravity) to directly generate quantum non-locality and entangled
states have not yet been studied.

In summary, the commutator $[\hat{x},\hat{p}]$ is the cornerstone
of the EPR argument, and the existence of minimal length (GUP)
replaces it with $[\hat{x},\hat{P}]$, a solid signal to reexamine
the EPR argument within the GUP framework. While an assumption in
QM is that a physical observable like momentum should be described
by an operator with real-valued eigenvalues, the canonical
momentum operator ($\hat{p}$) may take complex values
\cite{lit,22,Bonneau:1999zq,20,21,Uranga:2024ycr,Kim:2024bcx,Albert:2023kah}.
From a more general perspective, the existence of complex values
in quantum mechanics and their consequences and relationships with
quantum entanglement are longstanding fundamental topics
\cite{Albert:2023kah,AMJP,Renou:2021dvp,Li:2021uof,Chen:2021ril,Lancaster:2022vgi}.
As a result, and in the presence of minimal length, we intend to
investigate the possibility of $\hat{p}$ taking complex values and
also of creating quantum entanglement.

Therefore, if entangled states are found out, provided that they
$i$) are created only in the presence of minimum length, and $ii$)
include complex eigenvalues for $\hat{p}$, then the objectives
have been achieved. In the next section, the relationships between
the states (spaces) of $\hat{p}$ and $\hat{P}$ as well as the
possibility of producing complex eigenvalues for $\hat{p}$ are
investigated. After that, and equipped with the results of the
previous section, the ability of minimal length, leading to GUP,
in generating quantum non-locality is studied by providing
entangled states that also include complex quantum numbers. A
summary is presented in the last section.

\section{Canonical and generalized momentums, eigenstates and eigenvalues}

\textbf{In the purely quantum mechanical regime}, we basically
have $[\hat{x},\hat{p}]=i\hbar$ and

\begin{eqnarray}\label{2}
&&\hat{x}~|x_0\rangle=x_0~|x_0\rangle~\longrightarrow~\langle
x|x^\prime\rangle=\delta(x-x^\prime),\nonumber\\
&&\hat{p}~|p_0\rangle=p_0~|p_0\rangle~\longrightarrow~\langle p\
|p^\prime\rangle=\delta(p-p^\prime),\\
&&\int|p\rangle\langle p|~dp~=~\int|x\rangle\langle
x|~dx=\hat{1},\nonumber\\ &&\langle x|\hat{p}|p\rangle=p~\langle
x|p\rangle=-i\hbar\partial_x\langle x|p\rangle\Rightarrow\langle
x|p\rangle=\frac{\exp(\frac{ipx}{\hbar})}{\sqrt{2\pi\hbar}}.\nonumber
\end{eqnarray}

\noindent Therefore, $\exp(-\frac{ipx}{\hbar})$ corresponds to
state $|-p\rangle$ with eigenvalue $-p$. The coefficient
$\frac{1}{\sqrt{2\pi\hbar}}$ has been adopted to satisfy $\langle
p\ |p^\prime\rangle=\delta(p-p^\prime)$. $\hat{p}$ is Hermitian
for any two well-behaved wave functions $\psi$ and $\xi$ provided
that

\begin{equation}\label{wf}
\langle\psi|\hat{p}~\xi\rangle=\langle\hat{p}~\psi|\xi\rangle,
\end{equation}

\noindent valid whenever $\psi$ and $\xi$ go to zero as
$x\rightarrow\pm\infty$ \cite{lit}. More generally, $\hat{p}$ is
Hermitian if $\psi^*\xi|^b_a=0$, where $a$ and $b$ denote the
system boundary \cite{Kim:2024bcx}. Otherwise, complex eigenvalues
are dealt with
\cite{lit,22,Bonneau:1999zq,20,21,Uranga:2024ycr,Kim:2024bcx,Albert:2023kah}.

\textbf{In the presence of minimal length}, the generalized
momentum operator $\hat{P}(\equiv\hat{p}~(1+\beta\hat{p}^2))$
should be employed to measure momentum. It is also worthy to
remember that, based on quantum mechanics, the canonical momentum
operator ($\hat{p}$) is still a physical observable with its
Hilbert space ($|p\rangle$). In the minimal length regime, the
position operator remains unchanged (position representation
\cite{Brau:1999uv}). We have
$[\hat{x},\hat{P}]=i\hbar(1+3\beta\hat{P}^2)$ leading to
\cite{Hossenfelder:2012jw}

\begin{eqnarray}\label{7}
\Delta x\Delta P=\frac{\hbar}{2}(1+3\beta(\Delta P)^2),
\end{eqnarray}

\noindent that admits $\Delta
x_{min}=\sqrt{3\beta}~\hbar\equiv\sqrt{\beta_0}~l_p$ where $l_p$
denotes the Planck length, respectively
\cite{Hossenfelder:2012jw}. In this regime, we have
$\hat{H}=\frac{\hat{P}^2}{2m}$ for the Hamiltonian of a
non-relativistic free particle. Now, let us represent the
eigenstate of $\hat{P}$ with $|P\rangle$ and thus
$\hat{P}~|P_0\rangle=P_0~|P_0\rangle$. Moreover, since $\hat{P}$
and $\hat{p}$ commute, their info can be represented
simultaneously. But, it should be noted that since
$\hat{P}\neq\hat{p}$, their states (Hilbert spaces) are not the
same ($|p\rangle\neq|P\rangle$), a point easily verifiable using
$\langle x|P\rangle\neq\langle x|p\rangle$. Indeed,

\begin{eqnarray}\label{8}
\langle x|\hat{P}|P\rangle=P~\langle
x|P\rangle=-i\hbar\partial_x(1-\beta\hbar^2\partial_x^2)\langle
x|P\rangle,
\end{eqnarray}

\noindent whose solutions take the form

\begin{eqnarray}\label{9}
\exp(\frac{i\mathcal{P}x}{\hbar}),
\end{eqnarray}

\noindent with these values of $\mathcal{P}$

\begin{eqnarray}\label{10}
&&\mathcal{P}_1=\frac{\mathcal{A}}{6\beta}-\frac{2}{\mathcal{A}},\\ &&\mathcal{P}_2= -\frac{\mathcal{P}_1}{2}+i\frac{\sqrt{3}}{2}\bigg[\frac{\mathcal{P}_1}{2}+\frac{4}{\mathcal{A}}\bigg],\nonumber \\ &&\mathcal{P}_3=-\frac{\mathcal{P}_1}{2}-i\frac{\sqrt{3}}{2}\bigg[\frac{\mathcal{P}_1}{2}+\frac{4}{\mathcal{A}}\bigg],\nonumber\\
&&\mathcal{A}=\bigg(\big[108P+12\sqrt{3}\sqrt{\frac{27P^2\beta+4}{\beta}}\big]\beta^2\bigg)^{\frac{1}{3}}.\nonumber
\end{eqnarray}

Interestingly enough, it is seen that Eq.~(\ref{9}) is also the
eigenstates of the canonical momentum operator ($\hat{p}$) with
eigenvalues $\mathcal{P}_i$. Since $\beta$ is positive
\cite{Hossenfelder:2012jw,Kempf:1994su}, when
$\beta\rightarrow0^+$, we have $\mathcal{P}_1\rightarrow
P(\rightarrow p)$, $\mathcal{P}_2\rightarrow i\infty$, and
$\mathcal{P}_3\rightarrow-i\infty$. Additionally, $p~(1+\beta
p^2)=P$ is a third order equation of $p$ meaning that for each
eigenvalue $P$ there is not exactly one eigenvalue for $p$.
Indeed, there is a threefold degeneracy for each value of $P$
unveiled using $\hat{p}$.

%Therefore, in the presence of minimal length, $\mathcal{P}_2$ and $\mathcal{P}_3$ have also real part ($-\frac{\mathcal{P}_1}{2}$).

Here, the wave function~(\ref{9}) corresponding to $\mathcal{P}_2$
and $\mathcal{P}_3$ diverges as $x\rightarrow\infty$ or
$x\rightarrow-\infty$ (depending on the value of the imaginary
parts). Clearly, the condition $\xi^*\xi|^\infty_{-\infty}=0$
\cite{lit,Bonneau:1999zq,Kim:2024bcx} and thus Eq.~(\ref{wf}) are
not met, i.e. $\hat{p}$ is no longer Hermitian. Consequently, the
existence of minimal length may allow complex eigenvalues for
$\hat{p}$. It is finally useful to remember that, even in QM,
$\hat{p}$ may take complex values in various situations
\cite{lit,22,Bonneau:1999zq,20,21,Uranga:2024ycr,Kim:2024bcx,Albert:2023kah}
like whenever the integration interval does not obey the
translational symmetry \cite{Bonneau:1999zq} or when the
derivative of the wave function $\psi$ (or equally,
$\hat{p}~\psi$) is not normalizable \cite{lit}.

%Consequently, the existence of complex momentum in quantum
%mechanics may be a consequence of minimal length.

$|P\rangle$ and $|p\rangle$ denote the basis of the Hilbert spaces
corresponding to the operators $\hat{P}$ and $\hat{p}$,
respectively, and by considering the above argument, one can write
the general solution of Eq.~(\ref{8}) as

\begin{eqnarray}\label{22}
|P\rangle=\sum_{k=1}^3\alpha_k|p=\mathcal{P}_k\rangle,
\end{eqnarray}

\noindent where $\alpha_i$ denotes the expansion coefficients,
which brings us to

\begin{eqnarray}\label{23}
\langle x|P\rangle&=&\sum_{k=1}^3\alpha_k\langle
x|p=\mathcal{P}_k\rangle=\frac{1}{\sqrt{2\pi\hbar}}\sum_{k=1}^3\alpha_k\exp(\frac{i\mathcal{P}_kx}{\hbar}),\nonumber\\
\langle p|P\rangle&=&\int_{-\infty}^{\infty}\langle
p|x\rangle\langle x|P\rangle
dx\nonumber\\&=&\sum_{k=1}^3\alpha_k\big(\frac{1}{2\pi\hbar}\big)\int_{-\infty}^{\infty}\exp(-\frac{ipx}{\hbar})\exp(\frac{i\mathcal{P}_kx}{\hbar})
dx\nonumber\\&=&\sum_{k=1}^3\alpha_k\delta(p-\mathcal{P}_k).
\end{eqnarray}

\noindent Before proceeding further, it should be noted that
\cite{Kempf:1994su}

\begin{eqnarray}\label{231}
&&\langle P_1|P_2\rangle=\big(1+\beta P_1^2\big)\delta(P_1-P_2),\\
&&\hat{1}=\int|P\rangle\langle P|\frac{dP}{1+\beta P^2},\nonumber
\end{eqnarray}

\noindent and additionally, Eq.~(\ref{2}) is still valid for
operators $\hat{x}$ and $\hat{p}$, of course, in their spaces. As
the next step, since $dp=\frac{dP}{1+\beta P^2}$
\cite{Hossenfelder:2012jw}, one can write

\begin{eqnarray}\label{232}
&&\langle P_1|P_2\rangle=\sum_{k,m}\alpha_k^\ast\alpha_m\langle
p_k|p_m\rangle\Rightarrow\\
&&\int\langle P_1|P_2\rangle\frac{dP_2}{1+\beta
P_2^2}=\sum_{k,m}\int\alpha_k^\ast\alpha_m\langle p_k|p_m\rangle
dp_m,\nonumber
\end{eqnarray}

\noindent that leads to

\begin{eqnarray}\label{233}
\sum_{k=1}^3|\alpha_k|^2=1.
\end{eqnarray}

\noindent  Clearly, if measurement is done in the $|P\rangle$
bases, then we deal with only one state and thus one outcome. On
the other hand, if a purely quantum-mechanical apparatus is
employed (or equivalently, the measurement is performed in the
$|p\rangle$ space), then three outcomes $\mathcal{P}_i$ are likely
to occur after the collapse of the wave function.

%%%%%%%%%%%%%%%%%%%%%%%%%%%%%%%%%%%%%%%%%%%%%%%%%%%%%%%%%%%%%%%%%%%%%%%%
\section{Quantum entanglement production}

By coming back to Eqs.~(\ref{8}-\ref{10}), it is easily seen that
$-\mathcal{P}_k$ is a solution corresponding to state $|-P\rangle$
with eigenvalue $-P$ (just like as $\exp(-\frac{ipx}{\hbar})$ and
$-p$ in the purely quantum mechanical regime~(\ref{2})) and
interestingly enough, we have

\begin{eqnarray}\label{26}
-\mathcal{P}_k=\mathcal{P}_l+\mathcal{P}_m,
\end{eqnarray}

\noindent where $k,l,m\in\{1,2,3\}$. Comparing these results with
Eq.~(\ref{10}) for which

\begin{eqnarray}\label{27}
P=\mathcal{P}_j(1+\beta\mathcal{P}_j^2),
\end{eqnarray}

\noindent where $j=1,2,3$, one easily reaches

\begin{eqnarray}\label{28}
-\mathcal{P}_j(1+\beta\mathcal{P}_j^2)=-P,
\end{eqnarray}

\noindent yielding

\begin{eqnarray}\label{29}
|-P\rangle=\sum_{k=1}^3\gamma_k|-\mathcal{P}_k\rangle.
\end{eqnarray}

\noindent Therefore, $|-P\rangle$ consists of eigenstates of
canonical momentum with eigenvalues $-\mathcal{P}_j$.

At this step, consider a two-particle system whose total momentum
in the presence of minimal length (or equally, the total
generalized momentum) is zero so that $P_1=-P_2\equiv P$ and they
are moving along an arbitrary direction (such as the $x$ axis).
Clearly, the momentum state $|P,-P\rangle$ indicates that the
outcome of measuring $P_i$ is completely predictable. Now, bearing
Eqs.~(\ref{22}) and~(\ref{29}) in mind, one may generally write

\begin{eqnarray}\label{03}
|P,-P\rangle=\sum_{i,j=1}^{3}\alpha_i\gamma_j|\mathcal{P}_i,-\mathcal{P}_j\rangle,
\end{eqnarray}

\noindent apart from a normalization constant. Interestingly, the
generalized momentum conservation ($\sum_{j=1}^{j=2} P_j=0$) in
the $\hat{P}$-space also imposes the canonical momentum
conservation ($\sum_{j=1}^{j=2} p_j=$0) generating the $p_1=-p_2$
condition. Therefore, states do not meet the latter condition,
should be eliminated from the expression~(\ref{03}), done by
applying the projection operator
$\hat{\Pi}\big(\equiv\sum_i|\mathcal{P}_i\rangle_1\langle\mathcal{P}_i|\otimes|-\mathcal{P}_i\rangle_2\langle-\mathcal{P}_i|\big)$
on the state~(\ref{03}). Here, the indices $1$ and $2$ refer to
the first ($P$) and second ($-P$) particles, respectively, and one
gets the final state

\begin{eqnarray}\label{300}
\Pi|P,-P\rangle=\sum_{i=1}^{3}c_i|\mathcal{P}_i,-\mathcal{P}_i\rangle,
\end{eqnarray}%\equiv|\Omega\rangle

\noindent where $c_i\equiv\alpha_i\gamma_i$ and the state is
separable if only one $c_i$ is non-zero.

The probability of finding the state
$|\mathcal{P}_i,-\mathcal{P}_i\rangle$ is easily obtained as

\begin{eqnarray}\label{031}
\Gamma_i=\frac{|c_i|^2}{\sum_{k=1}^3|c_k|^2},
\end{eqnarray}

\noindent and the corresponding Von Neumann entropy reads as
follows \cite{VN}

\begin{eqnarray}\label{032}
S=-\sum_i\Gamma_i\ln\Gamma_i.
\end{eqnarray}

\noindent As long as at least two $c_i$ are non-zero, entropy is
greater than zero, revealing quantum non-locality. The Von Neumann
entropy also adopts its maximum whenever all states have the same
probability, i.e. $\Gamma_i=\frac{1}{3}$. Therefore, a novel
quantum non-locality seems to be unveiled that exposes itself in
the canonical momentum space.

%%%%%%%%%%%%%%%%%%%%%%%%%%%%%%%%%%%%%%%%%%%%%%%%%%%%%%
\section{Summary and Concluding Remarks}

Various theories studying the implications of gravity at quantum
scales propose a minimal length that generalizes momentum operator
and thus the uncertainty principle. Considering the first order of
generalization, an attempt to scrutinize the difference between
the $|P\rangle$ and $|p\rangle$ Hilbert spaces revealed a triple
degeneracy in the space $|p\rangle$ for each $|P\rangle$.
Interestingly, complex values for the canonical momentum seem
inevitable in the presence of a minimal length. Moreover,
considering a two-particle system whose total generalized momentum
is zero, a quantum entangled state has been discovered that may
reveal a new source of quantum non-locality.

%%%%%%%%%%%%%%%%%%%%%%%%%%%%%%%%%%%%%%%%%%%%%%%%%%%%%%
\section*{Acknowledgment}
The authors thank the respected referees for careful and
constructive comments. This work is based upon research funded by
Iran National Science Foundation (INSF) under project No. 4046949.
%%%%%%%%%%%%%%%%%%%%%%%%%%%%%%%%%%%%%%%%%%%%%%%%%%%%%%
\section*{Data availability}
%There is no data associated with this manuscript.
No data was used for the research described in the article.

%%%%%%%%%%%%%%%%%%%%%%%%%%%%%%%%%%%%%%%%%%%%%%%%%%%%%%
\section*{Declaration of competing interests}
The authors declare that they have no known competing financial
interests or personal relationships that could have appeared to
influence the work reported in this paper.

%%%%%%%%%%%%%%%%%%%%%%%%%%%%%%%%%%%%%%%%%%%%%%%%%%%%%%%

\end{document}